\documentclass[
reprint,
amsmath,amssymb,
aps,
pra,
superscriptaddress,
]{revtex4-2}

\usepackage{amsmath}
\usepackage{mathrsfs}
\usepackage{physics} 
\usepackage{xcolor}
\usepackage{afterpage}
\usepackage{adjustbox}
\usepackage{rotating}
\usepackage[sort&compress]{natbib}
\usepackage{graphicx}
\usepackage{dcolumn}
\usepackage{bm}
\usepackage[colorlinks=true, urlcolor=blue, pdfborder={0 0 0}]{hyperref}
\usepackage{braket}
\hypersetup{
linkcolor=blue,
citecolor=blue
}

\begin{document}

\preprint{APS/123-QED}

\title{Symmetry-Checking in Band Structure Calculations on a Noisy Quantum Computer}

\author{Shaobo Zhang}
\email{shaozhang@student.unimelb.edu.au}
\affiliation{School of Physics, The University of Melbourne, Parkville, 3010, Australia}

\author{Akib Karim}
\affiliation{Data61, CSIRO, 3168, Clayton Australia}

\author{Harry M. Quiney}
\affiliation{School of Physics, The University of Melbourne, Parkville, 3010, Australia}

\author{Muhammad Usman}
\affiliation{School of Physics, The University of Melbourne, Parkville, 3010, Australia}
\affiliation{Data61, CSIRO, 3168, Clayton Australia}

\date{\today}

\begin{abstract}

Band crossings in electronic band structures play an important role in determining the electronic, topological, and transport properties in solid-state systems, making them central to both condensed matter physics and materials science. The emergence of noisy intermediate-scale quantum (NISQ) processors has sparked great interest in developing quantum algorithms to compute band structure properties of materials. While significant research has been reported on computing ground state and excited state energy bands in the presence of noise that breaks the degeneracy, identifying the symmetry at crossing points using quantum computers is still an open question. In this work, we propose a method for identifying the symmetry of bands around crossings and anti-crossings in the band structure of bilayer graphene with two distinct configurations on a NISQ device. The method utilizes eigenstates at neighbouring $\mathbf{k}$ points on either side of the touching point to recover the local symmetry by implementing a character-checking quantum circuit that uses ancilla qubit measurements for a probabilistic test. We then evaluate the performance of our method under a depolarizing noise model, using four distinct matrix representations of symmetry operations to assess its robustness. Finally, we demonstrate the reliability of our method by correctly identifying the correct band crossings of AA-stacked bilayer graphene around $K$ point, using the character-checking circuit implemented on a noisy IBM quantum processor $ibm\_marrakesh$.
\end{abstract}

\maketitle


\section{\label{sec-Introduction}Introduction} 








Quantum computing has attracted significant attention due to its potential to offer exponential computational speedup in solving particular problems~\cite{daley2022practical}, such as random quantum circuit sampling~\cite{arute2019quantum, wu2021strong} and Gaussian boson sampling~\cite{deng2023gaussian}. However, today's quantum computers are not fault-tolerant and their performance is significantly limited by the presence of noise. These quantum processors are often referred to as noisy intermediate-scale quantum (NISQ) devices~\cite{preskill2018quantum}. Among the promising applications of NISQ devices is quantum chemistry, where parameterized quantum circuits are constructed to approximate the electronic structure, specifically, the energy spectra and eigenstates of quantum systems~\cite{peruzzo2014variational, cerezo2022variational, nakanishi2019subspace, parrish2019quantum, nam2020ground, ollitrault2020quantum, zhang2024full, karim2025fast, karim2024low}.

The band structure of a material plays a fundamental role in determining its electronic and optical properties and has been the focus of material science and condensed matter physics in the last century. It determines how electrons propagate through the solid and distinguishes between metals, semiconductors, and insulators. In particular, the band touching point, where two or more bands become degenerate or cross within the Brillouin zone, is of great interest since it leads to rich physical properties in quantum systems such as massless Dirac fermions and superconductivity in graphene~\cite{castro2009electronic, cao2018unconventional}, topologically protected surface states in topological insulators~\cite{hasan2010colloquium}, chiral anomaly and Fermi arcs in Weyl semimetals~\cite{armitage2018weyl}, and exceptional points in non-Hermitian physical systems~\cite{ashida2020non}. Such band crossing points are easily disrupted by noise on NISQ hardware, making it challenging to distinguish between true band crossings and anti-crossings. In other words, the identification of the touching points becomes difficult. Our focus here is to check that the symmetry at such a crossing point can be efficiently identified using a quantum computer.

In group theory, two energy bands are allowed to cross if they belong to different symmetry representations. Identifying band crossings is equivalent to determining whether the corresponding quantum states or symmetry-adapted linear combinations (SALC) of basis functions transform according to different irreducible representations (IRs). This can be verified by comparing their characters under a set of symmetry operations from the same point group. Since degenerate states always share the same symmetry representations, only non-degenerate states can be used to identify band crossings.

\begin{figure*}
\includegraphics[scale=0.2]{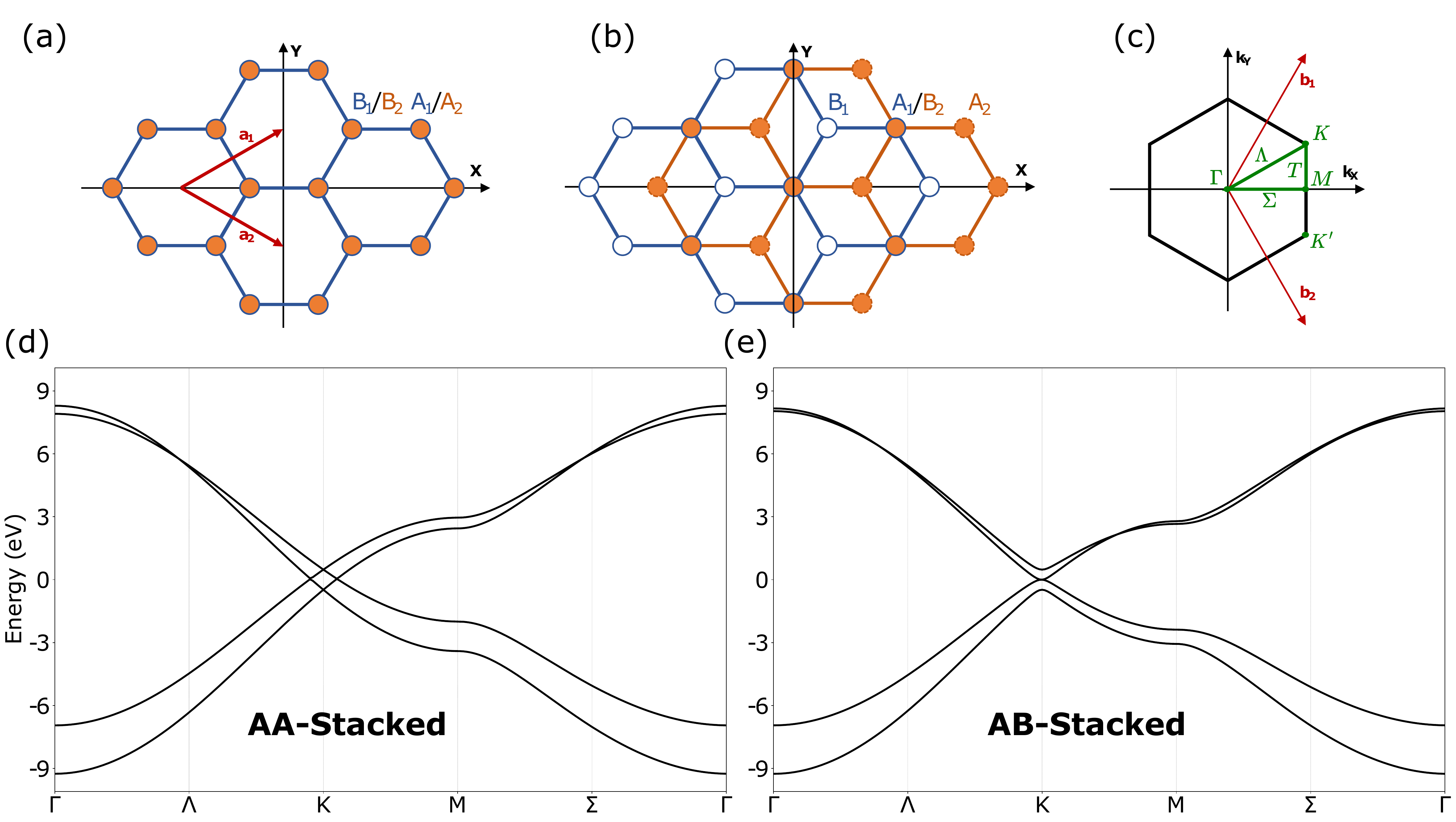}
\caption[Material structure]{
\label{fig1} 
The atomic and band structures of AA and AB-stacked bilayer graphene. (a) and (b) are the atomic configurations of AA and AB-stacked bilayer graphene, respectively. (c) is the corresponding Brillouin zone. (d) illustrates the band structure of the AA-stacked configuration, while (e) represents the band structure of the AB-stacked configuration.
}
\end{figure*} 

In this work, we propose a band ordering algorithm for identifying band crossings at critical $\mathbf{k}$ points in the band structure of the bilayer graphene system with two distinct configurations on a NISQ device. The algorithm leverages eigenstates at neighbouring $\mathbf{k}$ points on either side of the critical point to recover the local symmetry at the band touching points by employing a quantum circuit that uses ancilla qubit measurements for a probabilistic test. We demonstrate the reliability of our technique by accurately classifying band crossings in the band structure of bilayer Graphene on an IBM NISQ device accessed through cloud platform. The reported method can be adapted to identify band crossings and anti-crossings in a wide range of material systems. The rest of the paper is organized as follows. In Sec.~\ref{sec-Theory}, we present the theoretical foundation of our work. Specifically, the geometrical configurations of the bilayer graphene are described in Sec.~\ref{subsec-bi-graphene}; the associated tight-binding Hamiltonian is constructed in Sec.~\ref{subsec-tb-model}; the calculations of SALC basis functions are illustrated in Sec.~\ref{subsec-salc}; the quantum circuit used for band crossing check is introduced in Sec.~\ref{subsec-quantum-circ}. In Sec.~\ref{sec-Experiment}, we evaluate our method under two noisy scenarios: we first test our method on a depolarizing error noise simulator in Sec.~\ref{subsec-exp-depol}; we then test our method on an IBM NISQ device and show the experimental results in Sec.~\ref{subsec-exp-real-hardware}. The conclusions of our work are summarized in Sec.~\ref{sec-Conclusions}.


\section{\label{sec-Theory}Theory and  Methods}

\subsection{\label{subsec-bi-graphene}Bilayer Graphene}

In this article, we consider two flat graphene layers stacked on top of each other in two different configurations; the corresponding real space lattice structures are shown in Fig.~\ref{fig1}(a) and Fig.~\ref{fig1}(b), respectively. Each graphene layer consists of a honeycomb lattice with two carbon atoms labeled $A_i$ and $B_i$ in its unit cell, where 
$i$ denotes the layer index which can have the values of 1 and 2, representing the top and bottom layers, respectively. The in-plane nearest neighbour distance $a_{cc}$ between two carbon atoms $A_i$ and $B_i$ is around $1.42$\AA.

In the first configuration, the two graphene layers are aligned such that each atom in the top layer lies directly above the corresponding atom in the bottom layer. Specifically, atom $A_2$ in the bottom layer is directly beneath atom $A_1$ in the top layer, and atom $B_2$ is directly beneath atom $B_1$. This configuration is known as AA-stacked bilayer graphene, and the interlayer distance $d$ is reported as around $3.6\AA$~\cite{xu2010infrared,lobato2011multiple}. AA-stacked bilayer graphene possesses the same point group symmetry, $D_{6h}$, as monolayer graphene due to its real-space lattice alignment. This configuration belongs to a symmorphic space group and represents the highest possible symmetry achievable in a bilayer graphene system. The principal axis is defined as the axis that is perpendicular to the XY-plane, where the XY-plane lies midway between the two graphene layers. There are 24 symmetry operations classified into 12 classes in the point group.

In the second configuration, known as AB-stacked bilayer graphene, the two graphene layers are arranged such that only one sublattice of the top layer is positioned directly above a sublattice of the bottom layer. The interlayer distance $d$ is taken to be $3.3\AA$ in this work~\cite{xu2010infrared,tabert2012dynamical}. The AB-stacked bilayer graphene has 12 valid symmetry operations grouped into 6 classes, which is identified as the $D_{3d}$ point group symmetry. Since the $C_6$ operation is no longer valid in such a configuration, it has lower symmetry than AA-stacked bilayer graphene, therefore, it is a subgroup of the $D_{6h}$ point group.

In both configurations, each graphene layer retains the two-dimensional honeycomb lattice structure characteristic of a monolayer graphene, with the same primitive lattice vectors in the plane given by:
\begin{equation}
\label{eq-real-unit-vec}
\begin{aligned}
\mathbf{a_1} &= \frac{a_{cc}}{2} (3, \sqrt{3}, 0), \\
\mathbf{a_2} &= \frac{a_{cc}}{2} (3, -\sqrt{3}, 0).
\end{aligned}
\end{equation}
The three nearest neighbours of an atom $\alpha$ are connected by the vectors $\mathbf{\tau_i}$:
\begin{equation}
\label{eq-nearest-neighbour-vec}
\begin{aligned}
\mathbf{\tau}_1 &= a_{cc} (-1, 0, d), \\
\mathbf{\tau}_2 &= a_{cc} (\frac{1}{2}, \sqrt{3}, d), \\
\mathbf{\tau}_3 &= a_{cc} (\frac{1}{2}, -\sqrt{3}, d),
\end{aligned}
\end{equation}
where the value of $d$ depends on whether the nearest-neighbor interaction is in-plane or interlayer. The AA and AB-stacked bilayer graphene share the same Brillouin zone due to their identical in-plane lattice periodicity, which is illustrated in Fig.~\ref{fig1}(c). The reciprocal space lattice is defined by the unit vectors: 
\begin{equation}
\label{eq-recip-unit-vec}
\begin{aligned}
\mathbf{b_1} &= \frac{2\pi}{3a_{cc}} (1, \sqrt{3}, 0), \\
\mathbf{b_2} &= \frac{2\pi}{3a_{cc}} (1, -\sqrt{3}, 0).
\end{aligned}
\end{equation}
The hexagonal Brillouin zone has two inequivalent groups of $K$ points, $K = (2\pi/(3a_{cc}),2\pi/(3\sqrt{3}a_{cc}),0)$ and $K' = (2\pi/(3a_{cc}),-2\pi/(3\sqrt{3}a_{cc}),0)$, respectively.

\subsection{\label{subsec-tb-model}Tight-Binding Method}

To investigate the electronic structure of AA and AB-stacked bilayer graphene, we employ the tight-binding method to construct the Hamiltonian $H$. The $sp^2$ hybridization is assumed to be inert, hence, only the contribution from the $p_z$ atomic orbitals is considered. The unit cell of a bilayer graphene has four atoms, $A_1$ and $B_1$ on the top layer, and $A_2$ and $B_2$ on the bottom layer, thus, we can write the Bloch atomic orbital basis function in terms of the Fourier basis as:
\begin{equation}
\label{eq-bloch-func}
\begin{aligned}
\ket{\psi_\alpha(\mathbf{r_\alpha})} = \frac{1}{\sqrt{N}} \sum_{\mathbf{k}} e^{i\mathbf{k}\cdot\mathbf{r_\alpha}}\ket{u_{\alpha\mathbf{k}}(\mathbf{r_\alpha})},
\end{aligned}
\end{equation}
where $N$ is the number of unit cells in the quantum system, $\alpha \in \{A_1, B_1, A_2, B_2\}$ represents the atom, $\mathbf{k}$ is the coordinate in the Brillouin Zone, $\mathbf{r_\alpha}$ is the coordinate of atom $\alpha$ in the real space lattice. Considering only the nearest neighbour interaction, the corresponding tight-binding Hamiltonian can be written as: 
\begin{equation}
\label{eq-tb-ham}
\begin{aligned}
H_{tb} = \sum_{\alpha, \beta} \sum_{i=1}^3 h_{\alpha\beta} \ket{\psi_\alpha(\mathbf{r_\alpha})} \bra{\psi_\beta(\mathbf{r_\alpha}+\mathbf{\tau_i})},
\end{aligned}
\end{equation}
where $h_{\alpha\beta}$ is the $p_z$ orbital hopping integral between atoms on atomic site $\alpha$ and $\beta$, and $\alpha, \beta \in \{A_1, B_1, A_2, B_2\}$. Substitute Eq.~\ref{eq-bloch-func} into Eq.~\ref{eq-tb-ham}, the tight-binding Bloch Hamiltonian can be further expressed in matrix form as:
\begin{eqnarray}
\label{eq-tb-ham-k}
H(\mathbf{k})=
\begin{pmatrix}
0&f_1(\mathbf{k})&f_2(\mathbf{k})&f_3(\mathbf{k}) \\ f_1(\mathbf{k})^\ast&0&f_4(\mathbf{k})&f_2(\mathbf{k}) \\ f_2(\mathbf{k})^\ast&f_4(\mathbf{k})^\ast&0&f_1(\mathbf{k}) \\ f_3(\mathbf{k})^\ast&f_2(\mathbf{k})^\ast&f_1(\mathbf{k})^\ast&0 
\end{pmatrix},
\end{eqnarray}
where
\begin{subequations} \label{eq-hop}
\begin{align}
    f_1(\mathbf{k}) &= \sum_\mathbf{\tau} h_{A_1B_1}e^{i\mathbf{k}\cdot\mathbf{r}_{A_1B_1}}, \label{eq-hop1} \\
    f_2(\mathbf{k}) &= \sum_\mathbf{\tau} h_{A_1A_2}e^{i\mathbf{k}\cdot\mathbf{r}_{A_1A_2}}, \label{eq-hop2} \\
    f_3(\mathbf{k}) &= \sum_\mathbf{\tau} h_{A_1B_2}e^{i\mathbf{k}\cdot\mathbf{r}_{A_1B_2}}, \label{eq-hop3} \\
    f_4(\mathbf{k}) &= \sum_\mathbf{\tau} h_{B_1A_2}e^{i\mathbf{k}\cdot\mathbf{r}_{B_1A_2}}. \label{eq-hop4}
\end{align}
\end{subequations}
The statespace the Hamiltonian, $H_\mathbf{k}$, acts on is $\ket{\Psi_{p_z}} = (\ket{\psi_{A_1}},\ket{\psi_{B_1}},\ket{\psi_{A_2}},\ket{\psi_{B_2}})^T$. For instance, $f_1(\mathbf{k})$ calculates the hopping energy from atomic site $A_1$ to $B_1$, obtained by summing over all three in-plane nearest neighbors of $B_1$, denoted by the vectors $\mathbf{\tau}$; $f_2(\mathbf{k})$ calculates the hopping energy from atomic site $A_1$ to $A_2$, obtained by summing over all three interlayer nearest neighbors of $A_2$, denoted by the vectors $\mathbf{\tau}$. The hopping integral $h_{\alpha\beta}$ is calculated based on the model and empirical parameters presented in ~\cite{do2021proof, ho2020electronic, moon2012energy, moon2013optical, slater1954simplified, koshino2015interlayer},
\begin{equation}
\label{eq-hab}
\begin{aligned}
h_{\alpha\beta} 
=& 
V^0_{pp\sigma}\exp\left(\frac{d-r_{\alpha\beta}}{r_0}\right)\left(\frac{\mathbf{r}_{\alpha\beta} \cdot \mathbf{e}_z}{r_{\alpha\beta}}\right)^2 \\
&+
V^0_{pp\pi}\exp\left(\frac{a_{cc}-r_{\alpha\beta}}{r_0}\right)\left[1-\left(\frac{\mathbf{r}_{\alpha\beta} \cdot \mathbf{e}_z}{r_{\alpha\beta}}\right)^2\right], 
\end{aligned}
\end{equation}
where $V^0_{pp\sigma} \approx 0.48 $eV and $V^0_{pp\pi} \approx -0.27 $eV are known as the Slater-Koster parameters; $d$ is the interlayer distance between two graphene layers; $r_{\alpha\beta} = \abs{\mathbf{r}_{\alpha\beta}} = \abs{\mathbf{r_\alpha}-\mathbf{r_\beta}}$ is the distance between atomic sites $\alpha$ and $\beta$; $r_0 \approx 0.184\sqrt{3}a_{cc}$ describes the decay of the electronic hopping energy; $d$ is the interlayer distance between two graphene layers, which varies based on the different configuration; $\mathbf{e}_z = (0,0,1)$ is the unit vector in Z-axis.

The Hamiltonian $H(\mathbf{k})$ naturally satisfies the periodicity condition $H(\mathbf{k}+\mathbf{G}) = H(\mathbf{k})$, where $\mathbf{G} = n_1\mathbf{b}_1 + n_2\mathbf{b}_2$ denotes the reciprocal space lattice vector, and $n_1$ and $n_2$ are integers. In Fig.~\ref{fig1}(d) and Fig.~\ref{fig1}(e), we plot the band structure of AA and AB-stacked bilayer graphene, respectively, calculated by solving the tight-binding Hamiltonian introduced in this section, the calculation is performed along the $k$-path $\Gamma--K-M-\Gamma$ highlighted by the green path in Fig.~\ref{fig1}(c), with each segment of the path uniformly divided into 50 intervals.

\setlength{\tabcolsep}{12pt}
\begin{table*}[h]
\centering
\begin{tabular}{|ccccc|}
\hline
Configurations & $\mathbf{k}$ & Point Group & Irreps & SALC Basis Functions \\
\hline
AA-Stacked & $\Gamma$ & $D_{6h}$ & $A_{1g}$ & $(\ket{A_1}+\ket{B_1}+\ket{A_2}+\ket{B_2})/2$ \\
& & & $B_{2g}$ & $(\ket{A_1}-\ket{B_1}-\ket{A_2}+\ket{B_2})/2$ \\
& & & $A_{2u}$ & $(\ket{A_1}+\ket{B_1}-\ket{A_2}-\ket{B_2})/2$ \\
& & & $B_{1u}$ & $(\ket{A_1}-\ket{B_1}+\ket{A_2}-\ket{B_2})/2$ \\

& $\Lambda$ & $C_{2v}$ & $A_1$ & $(\ket{A_1}+\ket{B_1}-\ket{A_2}-\ket{B_2})/2$\\
& & & $A_2$ & $(\ket{A_1}+\ket{B_1}+\ket{A_2}+\ket{B_2})/2$\\
& & & $B_1$ & $(\ket{A_1}-\ket{B_1}-\ket{A_2}+\ket{B_2})/2$\\
& & & $B_2$ & $(\ket{A_1}-\ket{B_1}+\ket{A_2}-\ket{B_2})/2$\\

& $K$ & $D_{3h}$ & $E'$ & $(\ket{A_1}-\ket{A_2})/\sqrt{2}$, $(\ket{B_1}-\ket{B_2})/\sqrt{2}$\\
& & & $E''$ & $(\ket{A_1}+\ket{A_2})/\sqrt{2}$, $(\ket{B_1}+\ket{B_2})/\sqrt{2}$\\

& $T$ & $C_{2v}$ & $A_1$ & $(\ket{A_1}+e^{-i2\pi/3}\ket{B_1}+\ket{A_2}+e^{-i2\pi/3}\ket{B_2})/2$\\
& & & $A_2$ & $(\ket{A_1}-e^{-i2\pi/3}\ket{B_1}-\ket{A_2}+e^{-i2\pi/3}\ket{B_2})/2$\\
& & & $B_1$ & $(\ket{A_1}+e^{-i2\pi/3}\ket{B_1}-\ket{A_2}-e^{-i2\pi/3}\ket{B_2})/2$\\
& & & $B_2$ & $(\ket{A_1}-e^{-i2\pi/3}\ket{B_1}+\ket{A_2}-e^{-i2\pi/3}\ket{B_2})/2$\\

& $M$ & $D_{2h}$ & $A_{g}$ & $(\ket{A_1}+e^{-i2\pi/3}\ket{B_1}+\ket{A_2}+e^{-i2\pi/3}\ket{B_2})/2$\\
& & & $B_{3g}$ & $(\ket{A_1}-e^{-i2\pi/3}\ket{B_1}-\ket{A_2}+e^{-i2\pi/3}\ket{B_2})/2$\\
& & & $B_{1u}$ & $(\ket{A_1}+e^{-i2\pi/3}\ket{B_1}-\ket{A_2}-e^{-i2\pi/3}\ket{B_2})/2$\\
& & & $B_{2u}$ & $(\ket{A_1}-e^{-i2\pi/3}\ket{B_1}+\ket{A_2}-e^{-i2\pi/3}\ket{B_2})/2$\\

& $\Sigma$ & $C_{2v}$ & $A_1$ & $(\ket{A_1}+\ket{B_1}-\ket{A_2}-\ket{B_2})/2$\\
& & & $A_2$ & $(\ket{A_1}+\ket{B_1}+\ket{A_2}+\ket{B_2})/2$\\
& & & $B_1$ & $(\ket{A_1}-\ket{B_1}-\ket{A_2}+\ket{B_2})/2$\\
& & & $B_2$ & $(\ket{A_1}-\ket{B_1}+\ket{A_2}-\ket{B_2})/2$\\

\hline

AB-Stacked & $\Gamma$ & $D_{3d}$ & $A_{1g}$ & $(\ket{A_1}+\ket{B_2})/\sqrt{2}$ \\
& & & $A_{1g}$ & $(\ket{B_1}+\ket{A_2})/\sqrt{2}$ \\
& & & $A_{2u}$ & $(\ket{A_1}-\ket{B_2})/\sqrt{2}$ \\
& & & $A_{2u}$ & $(\ket{B_1}-\ket{A_2})/\sqrt{2}$ \\

& $\Lambda$ & $C_{2}$ & $A$ & $(\ket{A_1}+\ket{B_2})/\sqrt{2}$ \\
& & & $A$ & $(\ket{B_1}+\ket{A_2})/\sqrt{2}$ \\
& & & $B$ & $(\ket{A_1}-\ket{B_2})/\sqrt{2}$ \\
& & & $B$ & $(\ket{B_1}-\ket{A_2})/\sqrt{2}$ \\

& $K$ & $D_3$ & $A_1$ & $(\ket{A_1}-\ket{B_2})/\sqrt{2}$\\
& & & $A_2$ & $(\ket{A_1}+\ket{B_2})/\sqrt{2}$\\
& & & $E$ & $(\ket{A_2}-\ket{B_1})/\sqrt{2}$, $(\ket{A_2}+\ket{B_1})/\sqrt{2}$\\

& $T$ & $C_{2}$ & $A$ & $(\ket{A_1}+\ket{B_2})/\sqrt{2}$ \\
& & & $A$ & $(e^{-i4\pi/3}\ket{B_1}+\ket{A_2})/\sqrt{2}$ \\
& & & $B$ & $(\ket{A_1}-\ket{B_2})/\sqrt{2}$ \\
& & & $B$ & $(e^{-i4\pi/3}\ket{B_1}-\ket{A_2})/\sqrt{2}$ \\

& $M$ & $C_{2h}$ & $A_{g}$ & $(\ket{A_1}+\ket{B_2})/\sqrt{2}$ \\
& & & $A_{g}$ & $(e^{-i4\pi/3}\ket{B_1}+\ket{A_2})/\sqrt{2}$ \\
& & & $B_{u}$ & $(\ket{A_1}-\ket{B_2})/\sqrt{2}$ \\
& & & $B_{u}$ & $(e^{-i4\pi/3}\ket{B_1}-\ket{A_2})/\sqrt{2}$ \\

& $\Sigma$ & $C_s$ & $A'$ & $\ket{A_1}$ \\
& & & $A'$ & $\ket{B_1}$ \\
& & & $A'$ & $\ket{A_2}$ \\
& & & $A'$ & $\ket{B_2}$ \\

\hline
\end{tabular}
\caption{The symmetry-adapted linear combination (SALC) basis functions for AA- and AB-stacked bilayer graphene are analyzed at high-symmetry $\mathbf{k}$-points. For simplicity, we denote $\ket{\alpha} \equiv \ket{\psi_\alpha}$, where $\alpha \in {A_1, B_1, A_2, B_2}$ labels the atomic orbitals in the two layers. For $\mathbf{k}$-points lying between critical points, such as along the $\Lambda$ line connecting $\Gamma$ and $K$, the SALC basis functions are illustrated at the midpoint, e.g., $(\Gamma + K)/2$ is used to represent a representative $\mathbf{k}$-point on the $\Lambda$ path.}
\label{tab-salc-aa-ab}
\end{table*}

\begin{figure*}
\includegraphics[scale=0.2]{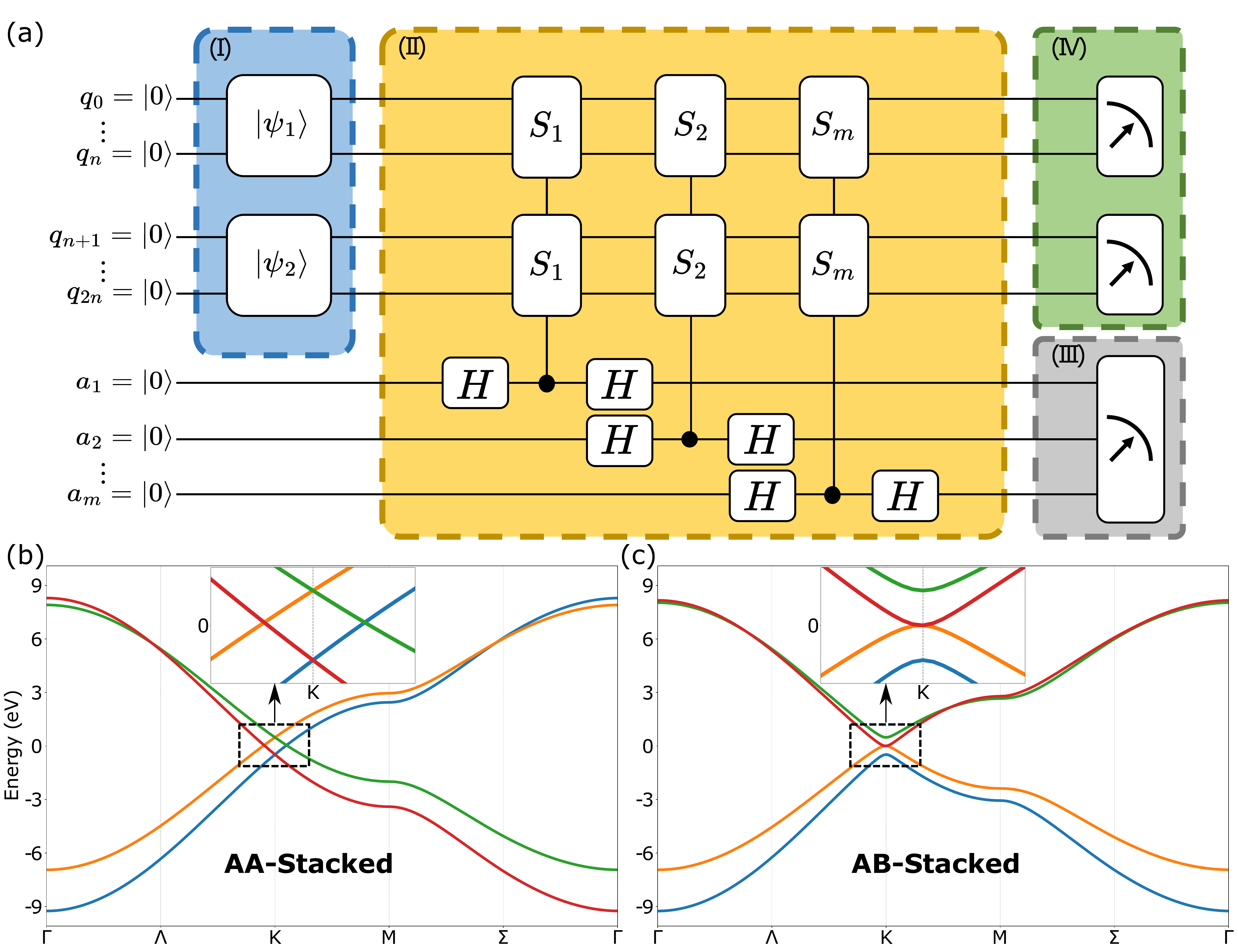}
\caption[Material structure]{
\label{fig2} 
The character check circuit and the corresponding numerical simulation results. (a) The character check circuit for checking if two quantum states $\ket{\psi_1}$ and $\ket{\psi_2}$ have the same character given the symmetry operation $S_1$ in the point group. The numerical simulations illustrating the band structures with band crossings for AA- and AB-stacked bilayer graphene are shown in (b) and (c), respectively.
}
\end{figure*} 

\subsection{\label{subsec-salc}Symmetry-Adapted Linear Combinations}

When $\mathbf{k} = 0$, the global phase is trivial, since $e^{i\mathbf{k}\cdot\mathbf{r}}=1$. The associated symmetry-adapted basis functions then only depend on the characters, $\chi$, from the corresponding character table of the point group symmetry and the atomic basis functions. Considering AA-stacked bilayer graphene as an example, the symmetry adapted linear combination of atomic orbital method can be summarized as follows: the reducible representation $\Gamma_{\sigma\pi}$ of the AA-stacked bilayer graphene is first constructed by applying 12 symmetry operations, $S$, that represent each class of the $D_{6h}$ point group. The corresponding matrix representations show the permutation of atoms, the character $\chi^S$ is given by the number of atoms left invariant under $S$. The reducible representation $\Gamma_{\sigma\pi}$ can then be further decomposed into the linear combination of IRs, the number of repetitions of $i$th IR $n_i$ is calculated by
\begin{equation}
\label{eq-rr-to-ir}
\begin{aligned}
n_i = \frac{1}{h}\sum_S(C^S\chi^S\chi^S_i),
\end{aligned}
\end{equation}
where $h$ is known as the order of the point group or the total number of symmetry operations in the point group, $C^S$ is the number of operations in the class, $\chi^S$ is the character of operation $S$ in $\Gamma_{\sigma\pi}$, and $\chi^S_i$ is the character of $S$ in the $i$th IR, therefore, the reducible representations of AA-stacked bilayer graphene can be decomposed as $\Gamma_{\sigma\pi} = A_{1g} \oplus B_{2g} \oplus A_{2u} \oplus B_{1u}$; finally, for $i$th IR found in the previous step, the symmetry-adapted basis function $\ket{\psi_i}$ can be obtained by
\begin{equation}
\label{eq-ir-to-sym-adapted}
\begin{aligned}
\ket{\psi_i} = \sum_{j}\chi_j^i\ket{\psi^i_{\alpha j}},
\end{aligned}
\end{equation}
where $\sum_j$ runs over all the 24 symmetry operations $j$ of the point group, $\chi^i_j$ is the character of $j$th symmetry operation given $i$th IR, and $\ket{\psi^i_{\alpha j}}$ is the atomic basis function. The normalized SALC basis functions of AA-stacked bilayer graphene at $\mathbf{k}=0$ are:
\begin{equation}
\label{eq-aa-gamma}
\begin{aligned}
\ket{\psi_{A_{1g}}} &= \frac{1}{2}(\ket{\psi_{A_1}}+\ket{\psi_{B_1}}+\ket{\psi_{A_2}}+\ket{\psi_{B_2}}), \\
\ket{\psi_{B_{2g}}} &= \frac{1}{2}(\ket{\psi_{A_1}}-\ket{\psi_{B_1}}-\ket{\psi_{A_2}}+\ket{\psi_{B_2}}), \\
\ket{\psi_{A_{2u}}} &= \frac{1}{2}(\ket{\psi_{A_1}}+\ket{\psi_{B_1}}-\ket{\psi_{A_2}}-\ket{\psi_{B_2}}), \\
\ket{\psi_{B_{1u}}} &= \frac{1}{2}(\ket{\psi_{A_1}}-\ket{\psi_{B_1}}+\ket{\psi_{A_2}}-\ket{\psi_{B_2}}).
\end{aligned}
\end{equation}

When $\mathbf{k} \neq 0$, the global phase factor associated with the Bloch states becomes significant when constructing symmetry-adapted basis functions. Not only does the point group symmetry of the quantum system vary with $\mathbf{k}$, but the principal axis may also change, such that two quantum states possessing the same point group symmetry at different $\mathbf{k}$ points can exhibit distinct symmetry properties. The corresponding principal axis can be found by applying the periodicity condition 
\begin{equation}
\label{eq-periodicity}
\begin{aligned}
    S\mathbf{k} = \mathbf{k} + \mathbf{G},
\end{aligned}
\end{equation}
where $S$ is the symmetry operation in the associated point group and $\mathbf{G}$ is the reciprocal space lattice vector. This affects the classification of states into IRs and is essential for accurately analyzing band structure symmetries and crossings. Given a Hamiltonian at a non-zero wavevector $\mathbf{k}$, which pocesses point group symmetry $\mathscr{S}$, the action of a symmetry operator $S \in \mathscr{S}$ on a Bloch function $\ket{\psi_\alpha(\mathbf{r_\alpha})} = e^{i\mathbf{k}\cdot\mathbf{r_\alpha}}\ket{u_{\alpha\mathbf{k}}(\mathbf{r_\alpha})}$ can be expressed as: 
\begin{equation}
\label{eq-extra-phase}
\begin{aligned}
S \ket{\psi_\alpha(\mathbf{r_\alpha})} 
&= \ket{\psi_\alpha(S^{-1}\mathbf{r_\alpha})} \\
&= e^{i\mathbf{k}\cdot(S^{-1}\mathbf{r_\alpha})}\ket{u_{\alpha\mathbf{k}}(\mathbf{S}^{-1}\mathbf{r_\alpha})} \\
&= e^{iS\mathbf{k}\cdot\mathbf{r_\alpha}}\ket{u_{\alpha\mathbf{k}}(S^{-1}\mathbf{r}_\alpha)} \\
&= e^{i(\mathbf{k} + \mathbf{G})\cdot\mathbf{r_\alpha}}\ket{u_{\alpha\mathbf{k}}(S^{-1}\mathbf{r}_\alpha)} \\
&= e^{i\mathbf{G}\cdot\mathbf{r_\alpha}}e^{i\mathbf{k}\cdot\mathbf{r_\alpha}}\ket{u_{\alpha\mathbf{k}}(S^{-1}\mathbf{r}_\alpha)} \\
&=
e^{i\mathbf{G}\cdot\mathbf{r_\alpha}}\ket{\psi_\alpha(\mathbf{r_\alpha})}, 
\end{aligned}
\end{equation}
where the extra phase is represented by $e^{i\mathbf{G}\cdot\mathbf{r}}$, and $\ket{\psi_\alpha(\mathbf{r_\alpha})}$ denotes the translational symmetry operation. The set of SALC functions at $\mathbf{k}$ points, corresponding to four different IRs, can be regarded as a valid solution of the Hamiltonian, as it diagonalizes the matrix representation.

Due to ambiguities in the definitions presented in Ref.~\cite{do2021proof}, we construct the SALC basis sets of AA- and AB-bilayer graphene that correctly block diagonalize the associated Hamiltonian to ensure consistent treatment. The corresponding SALC basis functions are presented in Table.~\ref{tab-salc-aa-ab}.

\begin{figure*}
\includegraphics[scale=0.204]{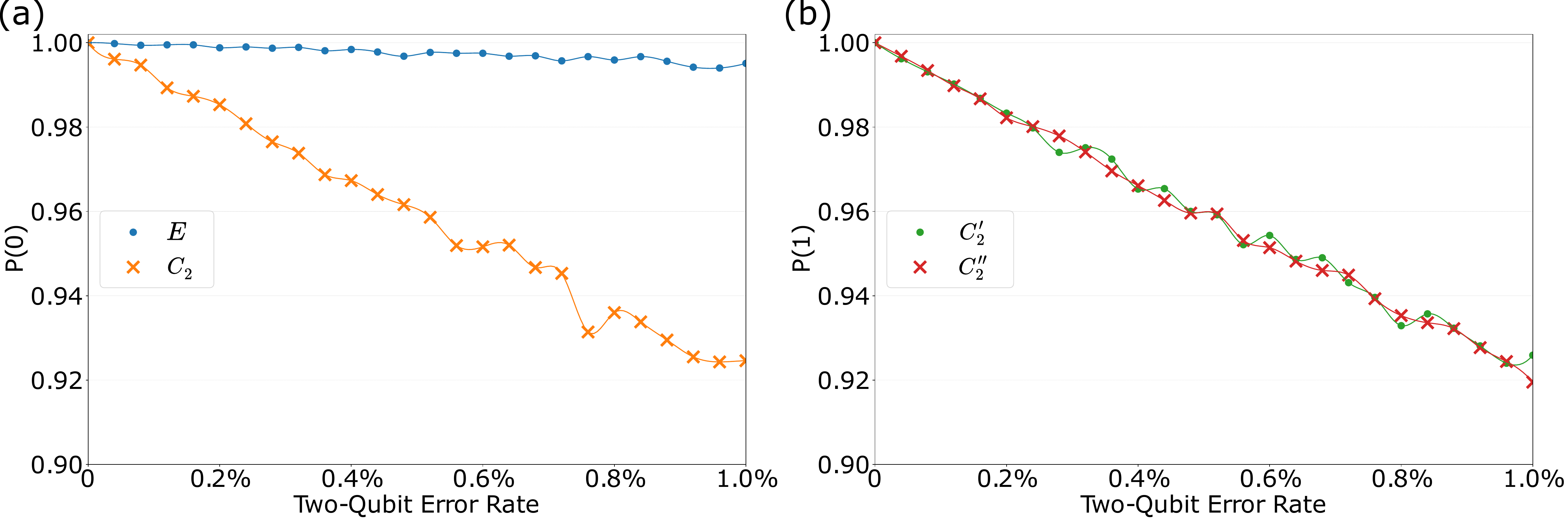}
\caption[Noisy simulation of RQEOM method]{
\label{fig_depol} 
Depolarizing noise model simulation of AA-stacked bilayer graphene at the $\Gamma$ point using the character-checking circuit for the ground and first excited states. Symmetry operations yielding the same characters for the two states are shown in (a), while those yielding different characters are shown in (b).
}
\end{figure*} 

\subsection{\label{subsec-quantum-circ}Character-Checking Circuit}

In quantum computing, given two quantum states $\ket{\psi_1}$ and $\ket{\psi_2}$, the swap test was developed to evaluate the similarity between two states~\cite{barenco1997stabilization}. It uses a single ancilla qubit, on which Hadamard and controlled-SWAP gates are applied. The probability of measuring $\ket{0}$ on the ancilla qubit is given by $1/2+\abs{\braket{\psi_1|\psi_2}}^2/2$, which implies that if two states are identical, only $\ket{0}$ will be measured; if two states are orthogonal, both $\ket{0}$ and $\ket{1}$ will be measured with $50\%$ probability. 

In this work, we generalize the swap test circuit for application in identifying if two bands belong to the same IR, as illustrated in Fig.~\ref{fig2}(a). Given a set of symmetry operations $S_{\ket{\psi_j}}^i$ belonging to the point group of quantum states $\ket{\psi_j}$ prepared on the target qubits in Fig.~\ref{fig2}(a)(I), the procedure for determining whether the two states share the same character under the symmetry operations $S_{\ket{\psi_j}}^i$ is shown in in Fig.~\ref{fig2}(a)(II). Taking the first symmetry operation $S_{\ket{\psi_j}}^1$ as an example, we analytically derive the ancilla qubit measurement outcome in Appendix~\ref{sec-threshold-finding}, with the resulting expression given as
\begin{equation}
\label{eq-swap-test}
\begin{aligned}
&H_{a_1} C_{a_1}S_{\ket{\psi_1}}^1S_{\ket{\psi_2}}^1H_{a_1}\ket{0_{a_1}}\ket{\psi_1}\ket{\psi_2} \\
=&
\frac{1}{2}[(1+c_1c_2)\ket{0}+(1-c_1c_2)\ket{1}]\ket{\psi_1}\ket{\psi_2},
\end{aligned}
\end{equation}
where $C_{a_1}S_{\ket{\psi_1}}^1S_{\ket{\psi_2}}^1$ denotes a Controlled-$S_{\ket{\psi_1}}^1$-$S_{\ket{\psi_2}}^1$ gate with the ancilla qubit $a_1$ as the control, applying two operations $S_{\ket{\psi_1}}^1$ and $S_{\ket{\psi_2}}^1$ on the target qubits where states $\ket{\psi_1}$ and $\ket{\psi_2}$ are prepared. Therefore, the measurement outcomes on the ancilla qubit correspond to $c_1$ and $c_2$ being the eigenvalues of the symmetry operator $S_{\ket{\psi_1}}^1\ket{\psi_1} = c_1\ket{\psi_1}$ and $S_{\ket{\psi_2}}^1\ket{\psi_2} = c_2\ket{\psi_2}$, respectively. In this work, we use the SALC basis functions listed in Table~\ref{tab-salc-aa-ab} as the eigenstates of the relevant symmetry operators for the experiments presented in Sec.~\ref{sec-Experiment}. Since only quantum states corresponding to one-dimensional IRs are used for band crossing checks, and valid symmetry operations will be applied to the quantum states, the associated characters can take values of either $+1$ or $-1$ and will be returned to the ancilla qubit as $c_1$ and $c_2$ for symmetry identification. If $c_1 = c_2 = \pm 1$, the probability of measuring state $\ket{0}$ on the ancilla qubit $a_1$ is $1$; if $c_1 = -c_2$, the probability of measuring state $\ket{1}$ on the ancilla qubit $a_1$ is $1$. A more detailed discussion is shown in Appendix~\ref{sec-threshold-finding}. To avoid phase mixing, the ancilla qubits should be measured in sequence rather than simultaneously. After the character-checking measurement in Fig.~\ref{fig2}(a)(III) is completed, the expectation value of quantum states with respect to the Hamiltonian, $\bra{\psi_i}H\ket{\psi_i}$, can be evaluated in Fig.~\ref{fig2}(a)(IV).


\section{\label{sec-Experiment}Results and Discussions}

In this section, we first analyze the performance of our method using AA-stacked bilayer graphene under a depolarizing error model on a classical quantum simulator, and the results are presented in Sec.\ref{subsec-exp-depol}. Subsequently, we confirm the validity of our method by conducting experiments on an IBM noisy quantum processor $ibm\_marrakesh$, and the results are discussed in Sec.\ref{subsec-exp-real-hardware}. The open-source package $qiskit$~\cite{qiskit} is used for the quantum computing part of our codes.

The algorithm can be summarized as follows. For each bilayer graphene configuration, the SALC basis functions and the relevant symmetry operations from the associated point group are first determined at high-symmetry $\mathbf{k}$ points. In classical calculations, the IRs are then assigned to the corresponding eigenstates. Band crossings can then be identified by examining whether the touching bands share the same symmetry: if they belong to different IRs, a crossing is allowed; otherwise, an anti-crossing is expected. In quantum computation, given two bands $\ket{\psi_i}$ and $\ket{\psi_j}$ prepared on the character-checking circuit, a set of symmetry operations from the relevant point group can be applied to determine their symmetry properties. If all ancilla qubits consistently yield measurement outcome $0$, the two states share the same IR, indicating an anti-crossing. In contrast, if the ancilla is in a superposition of $0$ and $1$, the states belong to different IRs, and a true band crossing is allowed. We present the numerical simulations of quantum computed band structure calculations of bilayer graphene with AA and AB-stacked configurations in Fig.~\ref{fig2}(b) and Fig.~\ref{fig2}(c), respectively. 

\subsection{\label{subsec-exp-depol}Depolarizing Noise Model}

In the presence of noise, the measurement outcomes of the ancilla qubits are contaminated with both outcomes $0$ and $1$, even when the prepared states are identical. To address this, we evaluate the robustness of our method in determining whether two states share the same IR under a uniform depolarizing noise model (details are in Appendix~\ref{sec-biased-depol-model}). Specifically, we examine the characters of the ground and first excited states of AA-stacked bilayer graphene at $\Gamma$ points by using the character-checking circuit. Both selected states belong to the $D_{6h}$ point group. The two-qubit error rate in the noise model varied from $0\%$ to $1\%$ based on the experiments reported~\cite{ali2024reducing, song2019quantum, li2023error}, which is shown as the x-axis in Fig.~\ref{fig_depol}, while the single-qubit error rate is set to be half of the two-qubit error rate.

\begin{figure*}
\includegraphics[scale=0.20]{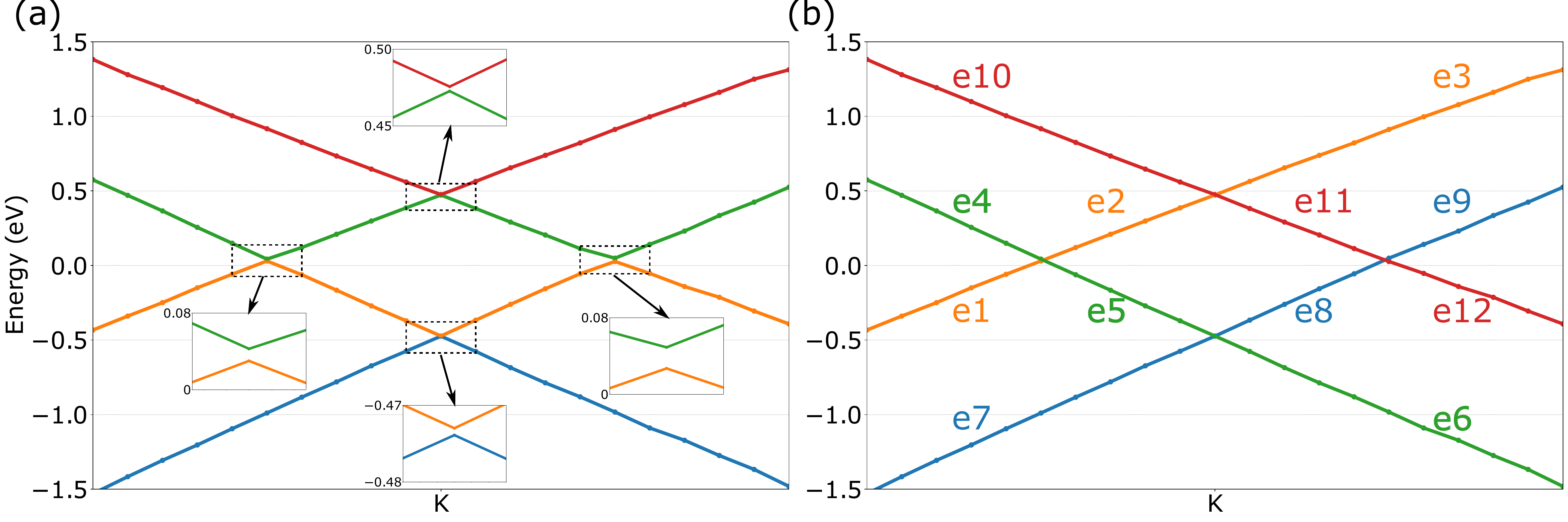}
\caption[Noisy simulation of RQEOM method]{
\label{fig3} 
The band crossing verification of AA-stacked bilayer graphene using the character-checking circuit on $ibm\_marrakesh$. The band structures are shown with raw data in (a), and with corrected band crossings verified by the character-checking circuit in (b). The eigenvectors used for verification are labeled from $e1$ to $e12$.
}
\end{figure*} 

We consider that the $D_{6h}$ point group has four distinct matrix representations, which can be represented by the identity operation $E$, three two-fold operations $C_2$, $C_2'$, and $C_2''$, respectively. The measurement results of $E$ and $C_2$ are illustrated in Fig.~\ref{fig_depol}(a), colored blue and orange, respectively. The y-axis corresponds to measuring $0$ on the ancilla qubit since both states have the same character under these two symmetry operations. As the error rate in the model increases, the blue line remains nearly constant, since the identity operation does not affect the quantum circuit. In contrast, the orange line decreases approximately linearly, reaching a probability of around $92.5\%$ for measuring $0$ at $1\%$, which is the practical error rate of two-qubit gates on quantum hardware. The measurement results of the other two matrix representations $C_2'$ (green) and $C_2''$ (red) are shown in Fig.~\ref{fig_depol}(b). The y-axis corresponds to measuring $1$ on the ancilla qubit since two states have different characters given the symmetry operations. Both lines decrease linearly with nearly identical slopes, reaching approximately $92.5\%$ at a $1\%$ two-qubit error rate. Since all four colored lines remain well above the $92\%$ under a realistic depolarizing error rate, we proceed to evaluate the performance of our method on actual quantum hardware in the following section.

\subsection{\label{subsec-exp-real-hardware}IBM Quantum Processor}

Although our method is applicable to both band crossings and anti-crossings, in this section, we focus on the AA-stacked bilayer graphene to demonstrate our method, as the AB-stacked configuration exhibits anti-crossings at the critical $K$ point, which are less informative for characterizing noise-affected behavior. Accordingly, we examine the band crossing of AA-stacked bilayer graphene in the region around the high-symmetry $K$ point (the inset of Fig.~\ref {fig2}(b)), demonstrating the experimental results in Fig.~\ref{fig3} by using the character-checking circuit on IBM's quantum hardware $ibm\_marrakesh$. The experiments are performed using qubits indexed 0, 1, 2, 3, and 4, highlighted in red in Fig.~\ref{fig-marakkesh}, and more details are in Appendix~\ref{sec-device-charac}. 

We first calculate the band structure of AA-stacked bilayer graphene on the noisy quantum processor by preparing the exact quantum state of the associated band, then measuring the expectation value based on the corresponding Hamiltonian $H(\mathbf{k})$. For each $\mathbf{k}$ point, we sort the calculated expectation value in an increasing order. We plot the data in Fig.~\ref{fig3}(a), where four insets illustrate band crossings that anti-cross each other due to the effect of noise. To investigate and restore the correct band crossings at these touching points, the selection of $\mathbf{k}$ points for character checking must ensure that they share the same point group symmetry at the $\mathbf{k}$ points before and after the touching point. Additionally, they should be sufficiently far from the touching point so that noise does not alter the energy ordering in its vicinity. Accordingly, we label each segment of bands from $e1$ to $e12$ respectively by following the selection rules, as shown in Fig.~\ref{fig3}(b). For each band touching point, we use the character-checking circuit to determine which pairs of bands share the same symmetry, based on their characters under a set of symmetry operations from the point group they belong to. Bands exhibiting identical characters for all tested operations are considered to have the same IR and are therefore labeled with the same color. For example, to determine the IRs of band segments $e5$, $e6$, $e7$, and $e8$, it is sufficient to check four pairs: ($e5$, $e6$), ($e5$, $e8$), ($e6$, $e7$), and ($e7$, $e8$). If the ancilla qubits yield all $0$s for the pairs ($e5$, $e6$) and ($e7$, $e8$), and all $1$s for the remaining pairs, with the measurement results staying above the confidence threshold, then we conclude that ($e5$, $e6$) and ($e7$, $e8$) share the same IRs. Such procedures can be done in parallel for efficiency. This indicates the presence of band crossings at the corresponding touching points. Therefore, we connect the lines at the $\mathbf{k}$ points used for character checking based on matching IRs. The intersection of two such lines indicates an approximate location of a band touching point. A similar approach can be used to identify avoided crossings. For example, consider a touching point surrounded by bands $e1$, $e2$, $e4$, and $e5$. If an avoided crossing is present, we first connect the line pairs ($e1$, $e5$) and ($e2$, $e4$), which correspond to different IRs, to approximate the location of the avoided crossing. We then reconnect the lines with the same IR through the touching point to correctly reconstruct the anti-crossing phenomenon. We show the band structure of AA-stacked bilayer graphene with restored band crossings in Fig.~\ref{fig3}(b) and present the hardware sampling results in Appendix~\ref{sec-samplings}. In this scenario, the point group symmetries of $\mathbf{k}$ points belong to $C_{2V}$ point group except $K$. It should be noted that, since the orientation of the principal axis varies as determined by the periodicity condition described in Eq.~\ref{eq-periodicity}, the symmetry operations need to be the representations of the associated point group. Therefore, a unitary transformation $U$ can be applied to the symmetry operation $S$ to unify the principal axis and phase for efficient circuit preparation. 


\section{\label{sec-Conclusions}Conclusions}

In this work, we present a method that uses a character-checking quantum circuit to accurately identify the symmetries of band crossings in the band structure of solid-state materials, and demonstrate through applications to AA- and AB-stacked bilayer graphene. We begin by correcting the inconsistencies in the SALC basis functions for both configurations at critical high-symmetry points in the Brillouin zone, as reported in Ref.~\cite{do2021proof}. We then evaluate the performance of our method under a depolarizing noise model using four distinct matrix representations of symmetry operations for the robustness testing. Finally, we apply our method to identify the correct band crossings in AA-stacked bilayer graphene, using the character-checking circuit implemented on a noisy quantum hardware $ibm\_marrakesh$. All sampling results stay above the established confidence threshold, enabling the successful restoration of the correct band crossing pattern in band structure calculation using quantum computers.


\begin{acknowledgments}

The research was supported by the University of Melbourne through the establishment of the IBM Quantum Network Hub at the University.

\end{acknowledgments}


\bibliography{main}


\appendix

\section{\label{sec-threshold-finding}Analysis Derivation}

In this appendix, we present the analytical derivation of the sampling results on the ancilla qubit $a_1$ of the character-checking circuit in the presence of noise. In this paper, given a symmetry operation $S_{\ket{\psi_j}}^i$ on two prepared quantum states $\ket{\psi_j}$ using the character-checking circuit, where $\ket{\psi_j}$ is a valid solution of $S_{\ket{\psi_j}}^i$, $S_{\ket{\psi_j}}^i\ket{\psi_j} = c_j\ket{\psi_j}$, $i$ denotes the $i$th symmetry operation in the corresponding point group and $j$ represents the $j$th prepared quantum state for checking. When $i=1$ and $j$ takes the value of $1$ and $2$, 
\begin{equation}
\label{eq-swap-test-details}
\begin{aligned}
&H_{a_1} C_{a_!}S_{\ket{\psi_1}}^1S_{\ket{\psi_2}}^1H_{a_1}\ket{0_{a_1}}\ket{\psi_1}\ket{\psi_2} \\
=&
\frac{1}{\sqrt{2}}H_{a_1} C_{a_1}S_{\ket{\psi_1}}^1S_{\ket{\psi_2}}^1(\ket{0_{a_1}}\ket{\psi_1}\ket{\psi_2}+\ket{1_{a_1}}\ket{\psi_1}\ket{\psi_2}) \\
=&
\frac{1}{\sqrt{2}}H_{a_1}(\ket{0_{a_1}}\ket{\psi_1}\ket{\psi_2}+\ket{1_{a_1}}S_{\ket{\psi_1}}^1\ket{\psi_1}S_{\ket{\psi_2}}^1\ket{\psi_2}) \\
=&
\frac{1}{\sqrt{2}}H_{a_1}(\ket{0_{a_1}}\ket{\psi_1}\ket{\psi_2}+\ket{1_{a_1}}c_1\ket{\psi_1}c_2\ket{\psi_2}) \\
=&
\frac{1}{2}[((\ket{0_{a_1}}+\ket{1_{a_1}})\ket{\psi_1}\ket{\psi_2}+ \\
&\quad c_1c_2(\ket{0_{a_1}}-\ket{1_{a_1}})\ket{\psi_1}\ket{\psi_2})] \\
=&
\frac{1}{2}[(1+c_1c_2)\ket{0}+(1-c_1c_2)\ket{1}]\ket{\psi_1}\ket{\psi_2}.
\end{aligned}
\end{equation}
Since $c_1$ and $c_2$ take the value of $\pm1$, the ancilla qubit would stay in $\ket{0}$ if $c_1=c_2$; and stay in $\ket{1}$ if $c_1=-c_2$. Therefore, the confidence threshold is set at $50\%$, as the two prepared states are eigenstates of the given symmetry operator. In this context, the majority of the measurement outcomes on the ancilla qubit are used to infer the character relationship between the two states. For example, if the ancilla qubit yields $40\%$ in the $\ket{0}$ and $60\%$ in the $\ket{1}$, it indicates that the two quantum states possess different characters under the applied symmetry operation.

Furthermore, when the prepared quantum states $\ket{\psi_k}$ and $\ket{\psi_l}$ are not eigenstates of a given symmetry operation, i.e., they do not satisfy $S^i_{\ket{\psi_j}}\ket{\psi_j} = c_j\ket{\psi_j}$ for $j = k, l$, but can instead be expressed as linear combinations of the eigenstates of the symmetry operator, we write $\ket{\psi_k} = \sum_m a_m \ket{\psi_m}$ and $\ket{\psi_l} = \sum_n b_n \ket{\psi_n}$, where $\ket{\psi_m}$ and $\ket{\psi_n}$ are the eigenstates of $S^i$ with corresponding eigenvalues $c_m$ and $c_n$. By implementing the character-checking circuit, we can determine whether $\ket{\psi_k}$ and $\ket{\psi_l}$ share the same symmetry with respect to the $i$th symmetry operator in the associated point group. 
\begin{equation}
\label{eq-swap-test-details-general}
\begin{aligned}
&H_{a_1}C_{a_1}S_{\ket{\psi_k}}^iS_{\ket{\psi_l}}^iH_{a_1}\ket{0_{a_1}}\ket{\psi_k}\ket{\psi_l} \\
=&
\frac{1}{\sqrt{2}}H_{a_1} C_{a_1}S_{\ket{\psi_k}}^iS_{\ket{\psi_l}}^i(\ket{0_{a_1}}\ket{\psi_k}\ket{\psi_l}+\ket{1_{a_1}}\ket{\psi_k}\ket{\psi_l}) \\
=&
\frac{1}{\sqrt{2}}H_{a_1}(\ket{0_{a_1}}\ket{\psi_k}\ket{\psi_l}+\ket{1_{a_1}}S_{\ket{\psi_k}}^i\ket{\psi_k}S_{\ket{\psi_l}}^i\ket{\psi_l}) \\
=&
\frac{1}{\sqrt{2}}H_{a_1}\sum_{m,n}a_mb_n(\ket{0_{a_1}}\ket{\psi_m}\ket{\psi_n}+ \\
&\qquad\qquad\qquad\quad\ \  \ket{1_{a_1}}S_{\ket{\psi_m}}^i\ket{\psi_m}S_{\ket{\psi_n}}^i\ket{\psi_n}) \\
=&
\frac{1}{\sqrt{2}}H_{a_1}\sum_{m,n}a_mb_n(\ket{0_{a_1}}\ket{\psi_m}\ket{\psi_n}+\ket{1_{a_1}}c_m\ket{\psi_m}c_n\ket{\psi_n}) \\
=&
\frac{1}{2}\sum_{m,n}a_mb_n[((\ket{0_{a_1}}+\ket{1_{a_1}})\ket{\psi_m}\ket{\psi_n}+ \\
&\qquad\qquad\quad c_mc_n(\ket{0_{a_1}}-\ket{1_{a_1}})\ket{\psi_m}\ket{\psi_n})] \\
=&
\frac{1}{2}\sum_{m,n}a_mb_n[(1+c_mc_n)\ket{0}+(1-c_mc_n)\ket{1}]\ket{\psi_k}\ket{\psi_l}.
\end{aligned}
\end{equation}
The probability of observing $\ket{0}$ and $\ket{1}$ on ancilla qubit are
\begin{equation}
\begin{aligned}
\label{eq-swap-test-details-general-results}
&P(0) = \frac{1}{4}\sum_{m,n}a_m^2b_n^2(1+c_mc_n)^2, \\
&P(1) = \frac{1}{4}\sum_{m,n}a_m^2b_n^2(1-c_mc_n)^2.
\end{aligned}
\end{equation}
This allows us to infer symmetry information even for approximated or noisy quantum states. In this scenario, finding the confidence threshold for the measurement results is determined by the product $c_m c_n$, representing the eigenvalues of the corresponding symmetry operator, and the product $a_m b_n$, representing the coefficients in the linear combinations of the two prepared states. In high-symmetry solid-state systems, half of the terms in Eq.~\ref{eq-swap-test-details-general-results} can be eliminated, enabling more efficient classical computation. For example, consider the $C_6$ rotation operation from the $D_{6h}$ point group used in this work. This operation cyclically permutes the atomic positions in AA-stacked bilayer graphene, and its matrix representation is given by:
\begin{equation}
\begin{aligned}
\label{eq-d6h-c6}
C_6 = 
\begin{pmatrix}
0 & 1 & 0 & 0\\
1 & 0 & 0 & 0\\
0 & 0 & 0 & 1\\
0 & 0 & 1 & 0\\
\end{pmatrix},
\end{aligned}
\end{equation}
and the corresponding eigenvalues $c_m$ and $c_n$ are $\{-1,-1,1,1\}$. The terms $1 + c_m c_n$ and $1 - c_m c_n$ in Eq.\ref{eq-swap-test-details-general-results} are nonzero only when $c_m$ and $c_n$ have the same or opposite signs, respectively. As a result, half of the terms vanish, which improves the efficiency of the calculation. Furthermore, Eq.\ref{eq-swap-test-details-general-results} indicates that the ancilla qubit generally yields both $\ket{0}$ and $\ket{1}$ outcomes, except in the special case where the prepared quantum states are eigenstates of the applied symmetry operator. This highlights the ongoing importance of accurately identifying the correct quantum eigenstates in quantum computing applications.

\section{\label{sec-biased-depol-model}Depolarizing Error Model}

The single-qubit depolarizing quantum channel is defined as
\begin{equation}
\begin{aligned}
\label{eq-biased-depol}
\mathcal{E}(\rho) = (1 - p) \rho + p\sum_ E r_E E^\dagger \rho E,
\end{aligned}
\end{equation}
where $p$ is the probability of a depolarizing error occurring on a single qubit, $E$ is either the single Pauli error channel or the summation of Pauli errors, $E \in \{X, Y, Z, I\}$, $r_\sigma$ defined as the proportion of a $\sigma$ error to all the noise. In this work, we simply take the uniform depolarizing error model, thus, $r_I = r_X = r_Y = r_Z = 0.25$.

\section{\label{sec-device-charac}Quantum Device Characteristics}

\begin{figure*}
\includegraphics[scale=0.5]{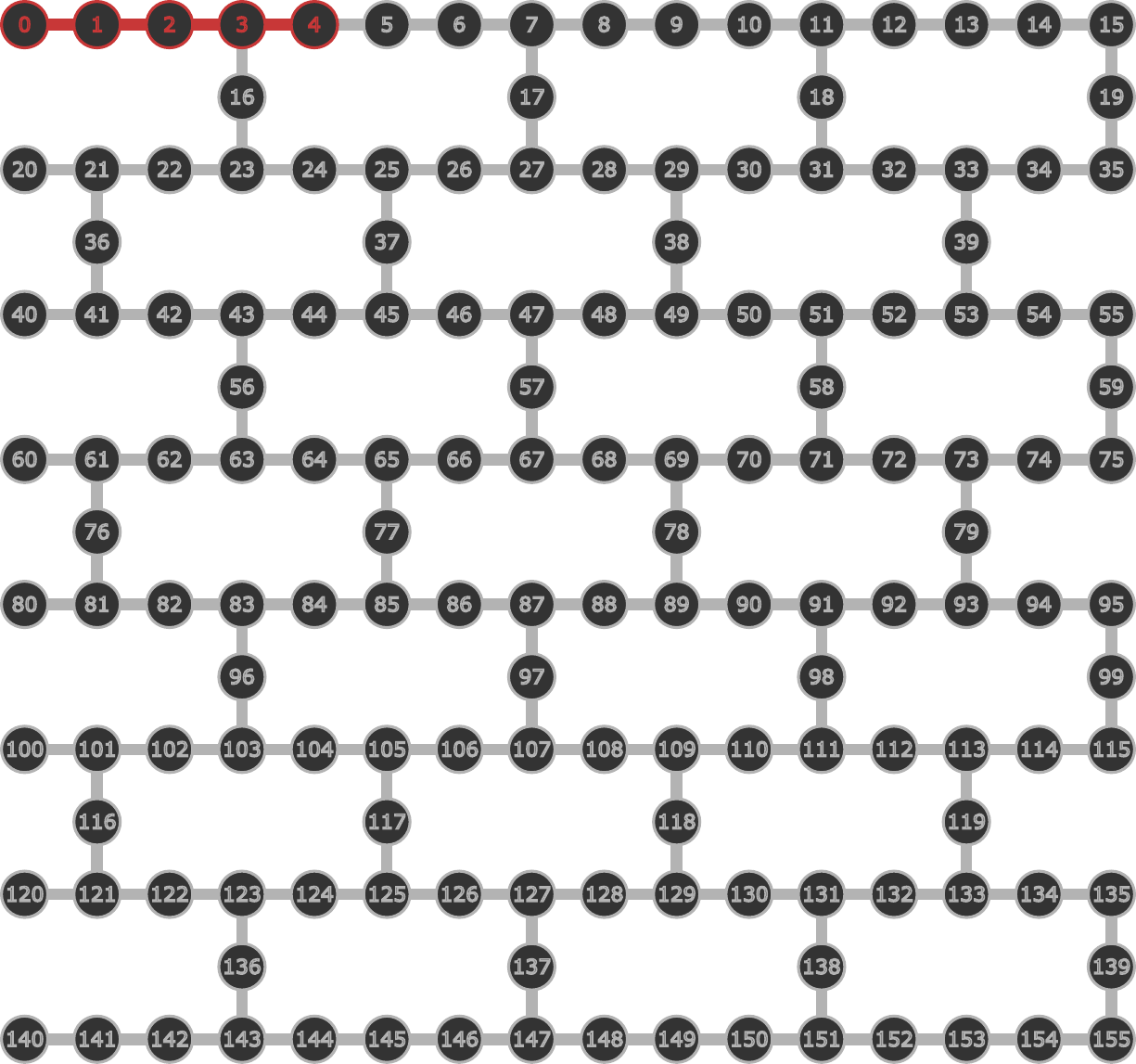}
\caption[Chip Geometry of Quantum Processor]{
\label{fig-marakkesh} 
The chip geometry of the IBM quantum processor $ibm\_marrakesh$. The qubits and connections used in this work are highlighted in red.
}
\end{figure*}

We illustrate the chip geometry of $ibm\_marrakesh$ in Fig.~\ref{fig-marakkesh}, where the qubits and connections highlighted red are used in our experiments. The corresponding noise spectroscopy of the highlighted qubits is shown in Table~\ref{tab-qubit-charac}, where the data was collected during the execution of our experiments in Section~\ref{subsec-exp-real-hardware}. 

\section{\label{sec-samplings}Hardware Sampling Results}

\begin{figure*}
\includegraphics[scale=0.204]{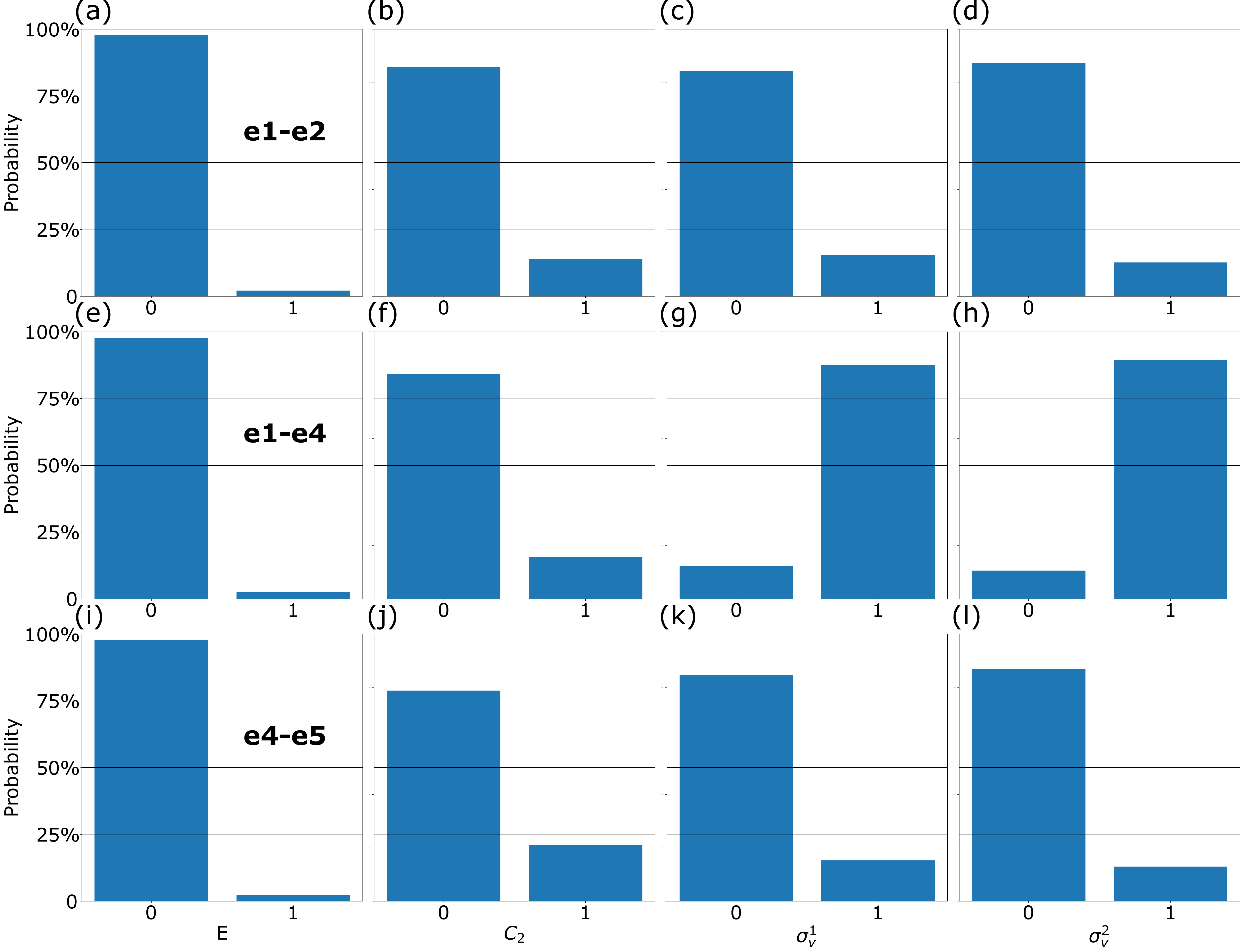}
\caption[Sampling results between e1, e2, e4 and e5]{
\label{fig3_sup1} 
The sampling results between the eigenvectors $e1$, $e2$, $e4$, and $e5$. Each column represents a distinct symmetry operation in the $C_{2v}$ point group, and each row corresponds to a pairwise comparison of two eigenvectors labeled in (a), (e), and (i), respectively. We take the ancilla qubit outcome corresponding to the state with a probability exceeding the threshold indicated by the solid black line.
}
\end{figure*}

\begin{figure*}
\includegraphics[scale=0.204]{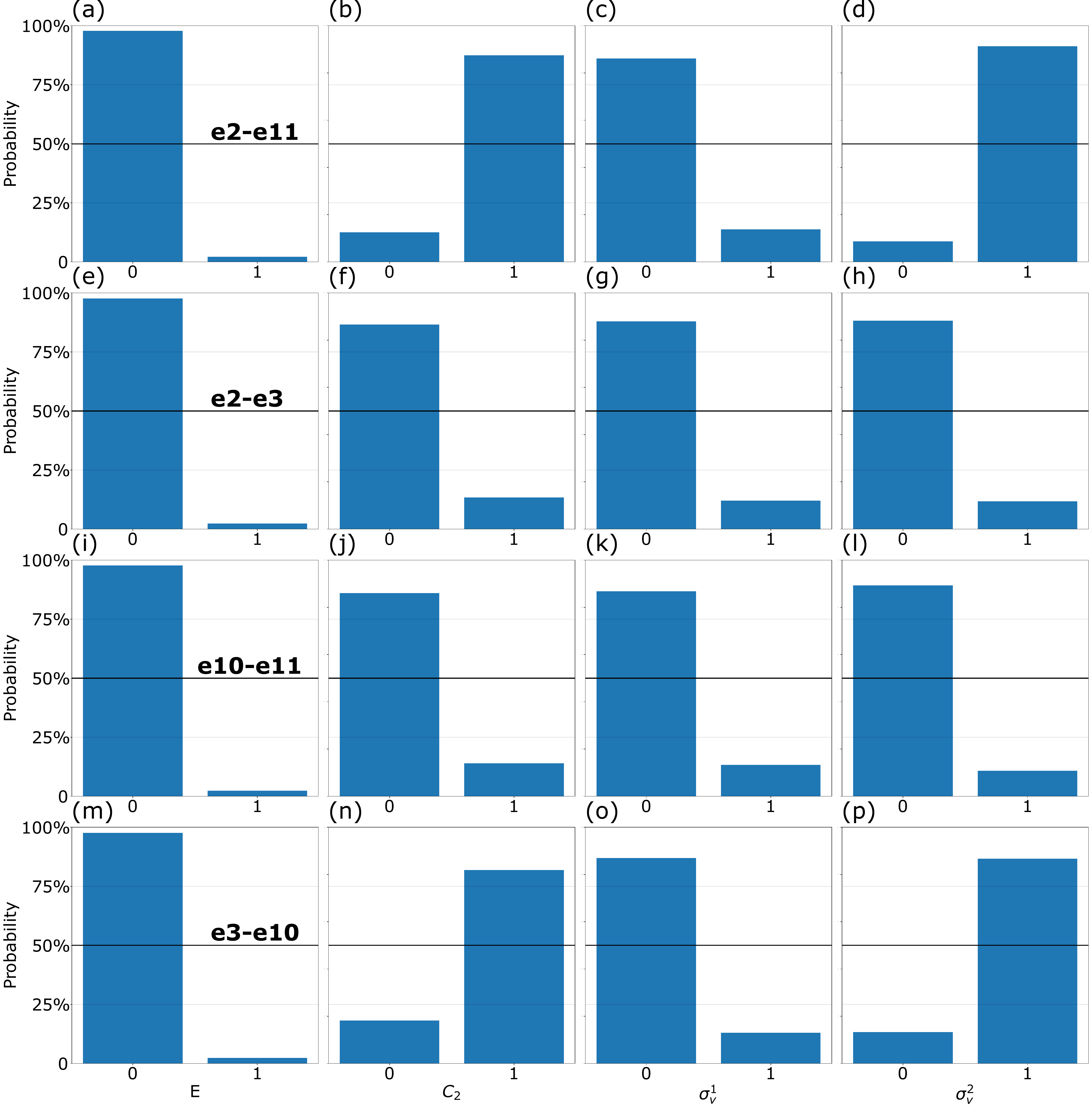}
\caption[Sampling results between e2, e3, e10 and e11]{
\label{fig3_sup2} 
The sampling results between the eigenvectors $e2$, $e3$, $e10$, and $e11$. Each column represents a distinct symmetry operation in the $C_{2v}$ point group, and each row corresponds to a pairwise comparison of two eigenvectors labeled in (a), (e), (i), and (m), respectively. We take the ancilla qubit outcome corresponding to the state with a probability exceeding the threshold indicated by the solid black line.
}
\end{figure*}

\begin{figure*}
\includegraphics[scale=0.204]{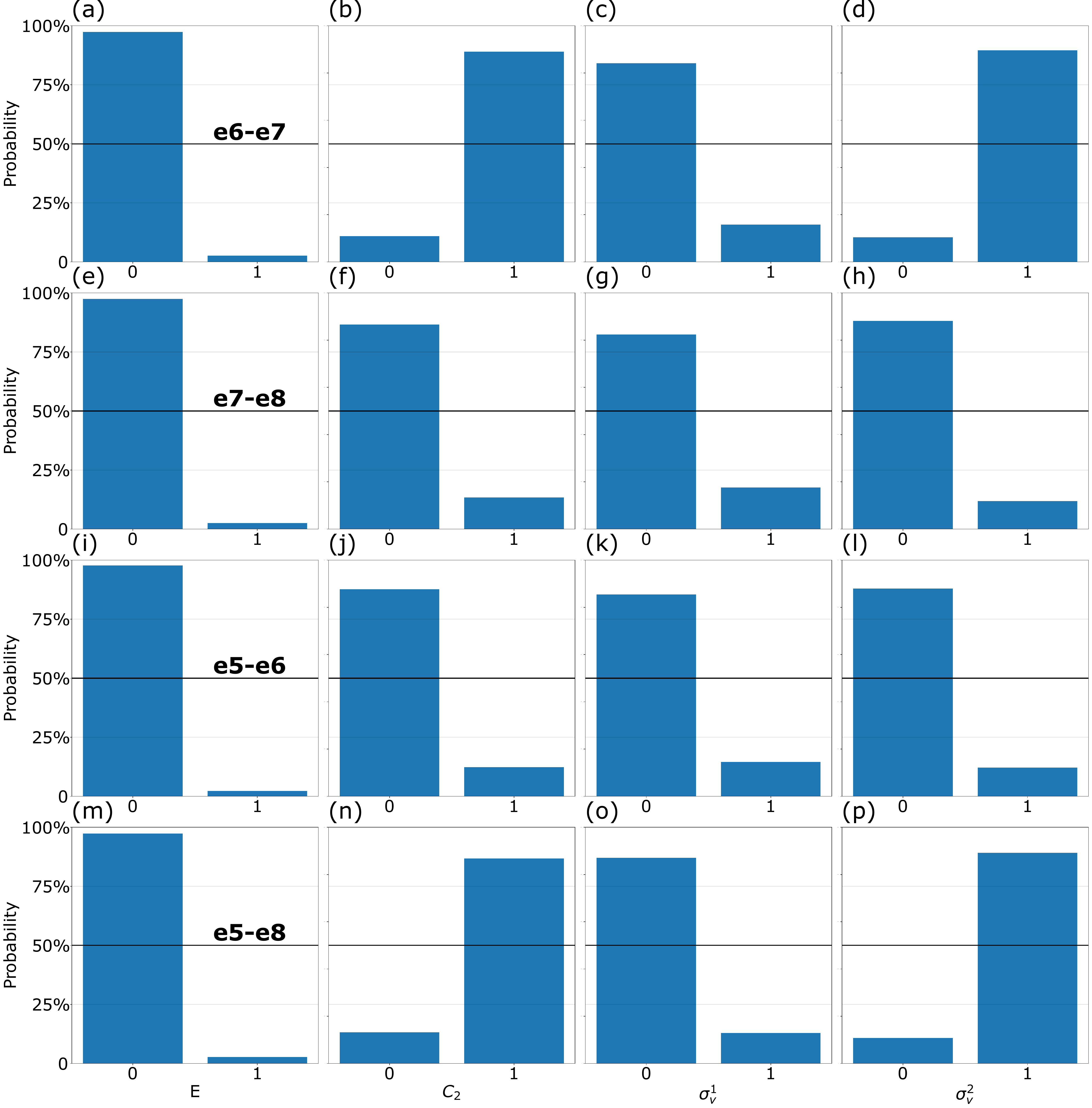}
\caption[Sampling results between e5, e6, e7 and e8]{
\label{fig3_sup3} 
The sampling results between the eigenvectors $e5$, $e6$, $e7$, and $e8$. Each column represents a distinct symmetry operation in the $C_{2v}$ point group, and each row corresponds to a pairwise comparison of two eigenvectors labeled in (a), (e), (i), and (m), respectively. We take the ancilla qubit outcome corresponding to the state with a probability exceeding the threshold indicated by the solid black line.
}
\end{figure*}

\begin{figure*}
\includegraphics[scale=0.204]{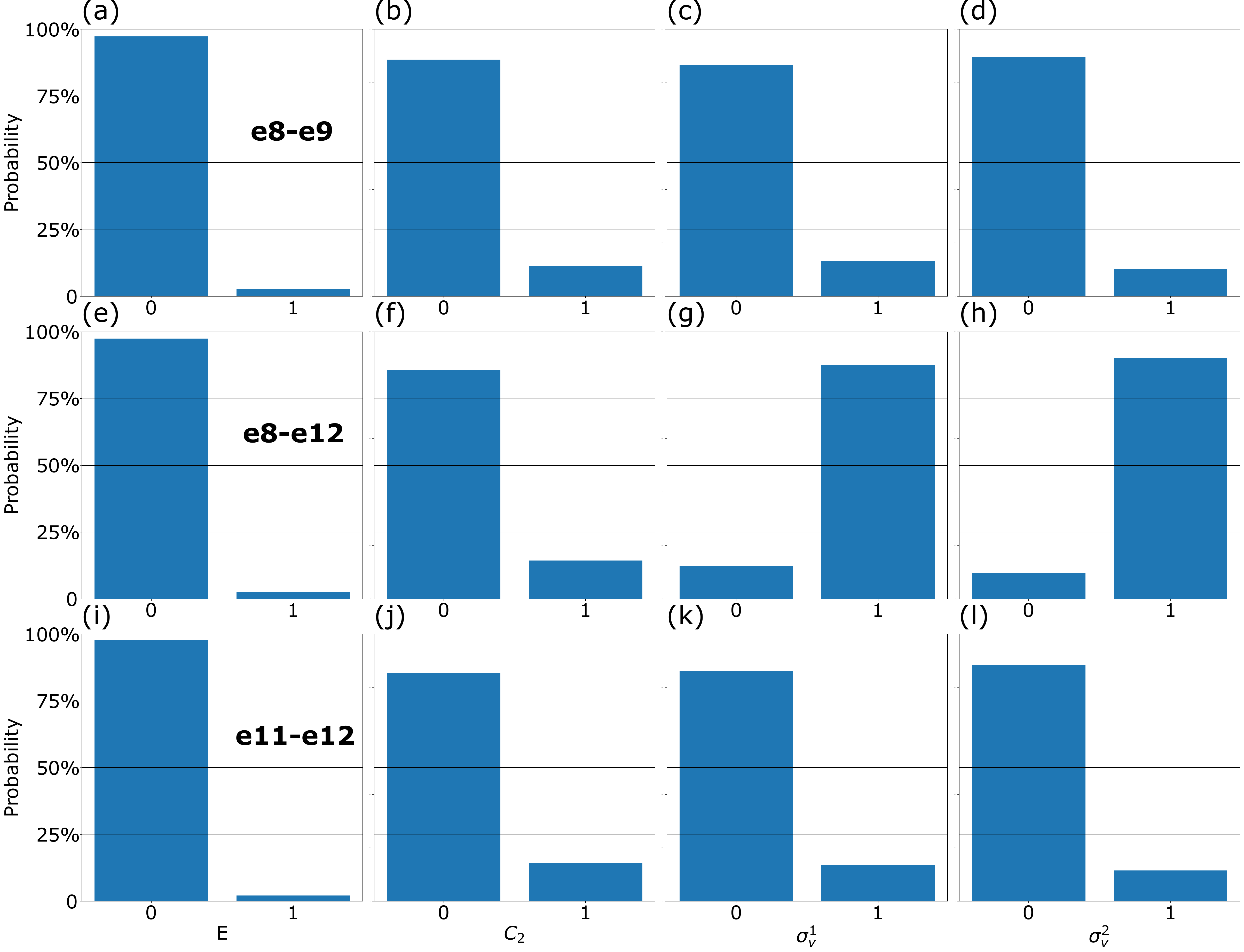}
\caption[Sampling results between e8, e9, e11 and e12]{
\label{fig3_sup4} 
The sampling results between the eigenvectors $e8$, $e9$, $e11$, and $e12$. Each column represents a distinct symmetry operation in the $C_{2v}$ point group, and each row corresponds to a pairwise comparison of two eigenvectors labeled in (a), (e), and (i), respectively. We take the ancilla qubit outcome corresponding to the state with a probability exceeding the threshold indicated by the solid black line.
}
\end{figure*}

We demonstrate the noisy quantum hardware sampling results for band crossing checking discussed Fig.~\ref{fig3}. The measurements of examining the crossing of bands $e1-e2$ and $e4-e5$ are shown in Fig.~\ref{fig3_sup1}; the results of bands $e2-e3$ and $e10-e11$ are illustrated in Fig.~\ref{fig3_sup2}; the results of bands $e5-e6$ and $e7-e8$ are represented in Fig.~\ref{fig3_sup3}; the results of bands $e8-e9$ and $e11-e12$ are demonstrated in Fig.~\ref{fig3_sup4}. All sampling results yield a dominant probability that stays above the confidence threshold indicated by the black horizontal line.

\setlength{\tabcolsep}{12pt}
\begin{table*}[h]
\centering
\begin{tabular}{|c|c|c|c|c|c|}
\hline
Qubit & 0 & 1 & 2 & 3 & 4 \\
\hline
$T_1$ ($\mu s$) & 410.55 & 178.27 & 171.96 & 335.39 & 246.15 \\
\hline
$T_2$ ($\mu s$) & 26.36 & 177.39 & 481.51 & 123.46 & 32.70 \\
\hline
Readout error (\%) & 0.34 & 0.59 & 2.71 & 4.27 & 8.76 \\
\hline
Prob meas0 prep1 (\%) & 0.44 & 1.07 & 2.10 & 3.42 & 17.43 \\
\hline
Prob meas1 prep0 (\%) & 0.24 & 0.10 & 3.37 & 5.13 & 0.10 \\
\hline
RX gate error (\%)& 0.01 & 0.03 & 0.01 & 0.02 & 0.73\\
\hline
$CZ_{q_0,q_n}$ error (\%)&  & 0.29 &  &  & \\
\hline
$CZ_{q_1,q_n}$ error (\%)& 0.29 &  & 0.16 &  & \\
\hline
$CZ_{q_2,q_n}$ error (\%)&  & 0.16 &  & 0.40 & \\
\hline
$CZ_{q_3,q_n}$ error (\%)&  &  & 0.40 &  & 3.56\\
\hline
$CZ_{q_4,q_n}$ error (\%)&  &  &  & 3.56 & \\
\hline

\hline
\end{tabular}
\caption{Device characteristics on $ibm\_marrakesh$ used in this work.}
\label{tab-qubit-charac}
\end{table*}  

\end{document}